\documentclass[12pt]{iopart}
\usepackage{graphicx,amsmath,bbm,mathrsfs,amssymb}
\newcommand{\bmr}{{\boldsymbol \varrho}_\kappa}
\newcommand{\bmrm}{{\boldsymbol \varrho}_M}
\newcommand{\bmn}{\boldsymbol \nu}
\begin{document}
\title[The quantum-classical transition in thermally seeded PDC]{The
quantum-classical transition in thermally seeded parametric downconversion}
\author{Ivo P Degiovanni$^{1}\footnote{i.degiovanni@inrim.it}$, Marco
Genovese$^1$, Valentina Schettini$^1$, Maria Bondani$^2$, Alessandra
Andreoni$^{3,4}$, Matteo G A Paris$^{5,6,7}$}
\address{$^1$ Istituto Nazionale di Ricerca Metrologica,
I-10135, Torino, Italia}
\address{$^2$ National Laboratory for Ultrafast and Ultraintense
Optical Science - C.N.R.-I.N.F.M., Como, Italia}
\address{$^3$ Dipartimento di Fisica e Matematica, Universit\`a
dell'Insubria, I-22100, Como, Italia}
\address{$^4$ C.N.I.S.M., U.d.R. Como, I-22100, Como, Italia}
\address{$^5$ Dipartimento di Fisica, Universit\`a di Milano, I-20133 Milano, Italia}
\address{$^6$ C.N.I.S.M., U.d.R Milano Universit\`a, I-20133, Milano, Italia}
\address{$^7$ I.S.I. Foundation, I-10133 Torino, Italia}
\begin{abstract}
We address the pair of conjugated field modes obtained from
parametric-downconversion as a convenient system to analyze the
quantum-classical transition in the continuous variable regime.  We
explicitly evaluate intensity correlations, negativity and
entanglement for the system in a thermal state and show that a
hierarchy of nonclassicality thresholds naturally emerges in terms
of thermal and downconversion photon number.  We show that the
transition from quantum to classical regime may be tuned by
controlling the intensities of the seeds and detected by intensity
measurements.  Besides, we show that the thresholds are not affected
by losses, which only modify the amount of nonclassicality. The
multimode case is also analyzed in some detail.
\end{abstract}
\maketitle
\section{Introduction}
The boundary between quantum and classical physics has been
controversial \cite{wz91,wz03,sch03,sch06,gen05,R1,R2,R3,R4} ever
since the early days of quantum mechanics. Nevertheless, the
solution to this problem is very important for several fundamental
issues in quantum and atomic optics and, more generally, in quantum
measurement theory \cite{hab98, bha00}. More recently, with the
development of quantum technology, the issue gained new interest
since nonclassical features, in particular entanglement, represent
practical resources to improve quantum information processing.
\par
As a matter of fact, quantum decoherence, \textit{i.e.} the
dynamical suppression of quantum interference effects, cannot be the
unique criterion to define a classical limit \cite{sc05}, which
should emerge from an operational approach suitably linked to
measurement schemes \cite{si04,si05,sa05,si03, R5, R6, R7, R8, R9,
R10, R11, R12, R13, R14, R15, R16, R17}. To this aim here we address
the bipartite system made by the pair of field modes obtained from
parametric downconversion (PDC) as a convenient physical system to
analyze the quantum-classical transition in the continuous variable
regime.  We consider a PDC process seeded by thermal radiation, a
scheme that we already investigated in ghost-imaging/diffraction
experiments \cite{Pud07,DBA07,BPD08} and where it has been shown
that both entanglement and intensity correlations may be tuned upon
changing the intensities of the seeds \cite{DBA07}.  This, in turn,
puts forward the PDC output as a natural candidate to investigate
the quantum-classical transitions in an experimentally feasible
configuration.  We focus on some relevant parameters employed to
point out the appearance of quantum features, namely sub-shot-noise
correlations, negativity and entanglement. We analyze at varying the
mean photon numbers of the interacting fields the different
nonclassicality thresholds that appear. Remarkably, the
corresponding transitions from classical to quantum domain may be
observed experimentally by means of intensity measurements, thus
avoiding full state reconstruction by homodyne or other
phase-sensitive techniques \cite{raymerLNP, darLNP}.
\par
The paper is structured as follows. In the next section we review the
PDC process, establish notation and introduce the nonclassicality
parameters we are going to analyze. In Section 3 we analyze the effect
of losses, whereas in Section 4 we discuss the generalization of our
analysis to the multimode case. Finally, Section 5 closes the paper with
some concluding remarks.
\section{Parametric downconversion with thermal seeds}
\label{s:MMTPDC} The evolution of a pair of field modes under PDC is
described by the unitary operator ${U}_\kappa=\exp(i \kappa
{a_1}{a_2} + h.c.)$, where $\kappa$ is the coupling constant and
$a_j$ are the mode operators $(j=1,2)$. In the following we consider
the two modes initially in a thermal state, {\em i.e.} excited in a
factorized thermal state $\bmn=\nu_1 \otimes \nu_2$,
$\nu_j=\sum_{n=0}^{\infty} p_{j}(n)~ |n\rangle_{jj}\!\langle n|$
being a single-mode thermal state with $\mu_j$ mean number of
photons, {\em i.e.} $p_{j}(n)=\mu_{j}^{n} (1+\mu_{j})^{-(n+1)}$. The
density matrix at the output is given by ${\bmr} =
{U}_\kappa\:\bmn\:{U}^{\dagger}_\kappa$, whereas the output modes
are given by $A_j={U}_{\kappa}^{\dag}{a}_{j}{U}_{\kappa}=\alpha a_j
+ e^{i\varphi}\beta a_{j'}^\dag$ (with $j=1,2$ and $j\neq j'$) where
$\alpha= \cosh| \kappa|$, $\beta = \sinh |\kappa |$ and $\varphi$ is
the coupling ({\em i.e.} pump) phase. The statistics of the two
output modes, taken separately, are those of a thermal state
\cite{DBA07}, {\em i.e.} $\langle {n}_{j} \rangle =\mu_{j}+
\mu_{\kappa}(1+ \mu_{1}+ \mu_{2})$, $\langle \Delta n_{j}^{2}\rangle
= \langle {n}_{j} \rangle (\langle {n}_{j} \rangle+1)$, where
$n_{j}= a_{j}^{\dag} a_{j} $ and
$\mu_\mathrm{\kappa}=\sinh^2|\kappa|$ is the mean number of photons
due to spontaneous PDC; the symbols $\langle ... \rangle$ and
$\Delta$ denote $\langle {O} \rangle= \mathrm{Tr}[{O}{\varrho}]$ and
$\Delta{O}= {O}- \langle {O} \rangle$, respectively. Notice that the
case of vacuum inputs, $\bmn= |0\rangle\langle
0|_{1}\otimes|0\rangle\langle 0|_{2}$, corresponds to spontaneous
downconversion, {\em i.e.} to the generation of the so-called pure
twin-beam state (TWB) $|{\boldsymbol \psi}_\kappa\rangle\rangle =
U_\kappa |0\rangle=\alpha^{-1} \sum_n (\beta/\alpha)^n\:
|n\rangle\otimes|n\rangle$ \cite{AK90}.
\subsection{Intensity correlations}
The shot-noise limit (SNL) in a photodetection process is defined as
the lowest level of noise that can be achieved by using
semiclassical states of light \cite{SNL1,SNL2, SNL3}, that is
Glauber coherent states. On the other hand, when a noise level below
the SNL is observed, we have a genuine nonclassical effect.  For a
two-mode system if one measures the photon number of the two beams
and evaluates the difference photocurrent $H = n_{1}-n_{2}$ the SNL
is the lower bound to the fluctuations $\langle \Delta H^{2}\rangle$
that is achievable with classically coherent beams, {\em i.e.}
$\langle \Delta H^{2}\rangle = \langle {n}_{1} \rangle +  \langle
{n}_{2}\rangle$.
\par
Let us consider a simple measurement scheme where the modes at the output of the PDC crystal
are individually measured by direct detection and the resulting
photocurrents are subtracted from each other to
build the difference photocurrent. We have quantum noise
reduction, {\em i.e.} violation of the SNL, when
$\langle \Delta H^{2} \rangle <  \langle {n}_{1} \rangle + \langle
{n}_{2} \rangle$, that is \cite{DBA07}
\begin{equation}
\mu_1^2 + \mu_2^2 < 2 \mu_\kappa (1+ \mu_1 + \mu_2)
\label{ssn}
\end{equation}
In order to quantify intensity correlations and to evaluate the amount of
violation of the SNL we introduce the parameter
\begin{equation} \label{Pssn0}
\gamma_c= 1- \frac{\langle \Delta H^{2}\rangle}{\langle {n}_{1}
\rangle + \langle {n}_{2} \rangle} .
\end{equation}
The value $\gamma_c= 0$ corresponds to noise at the SNL, whereas the presence
of nonclassical intensity correlations leads to $0 < \gamma_c \leq
1$. For the pair of modes at the output of the PDC crystal we obtain
\begin{equation} \label{Pssn}
\gamma_c= \frac{2 ~ \mu_\kappa (1+ \mu_1 + \mu_2)- \mu_1^2 -
\mu_2^2}{2 ~ \mu_\kappa (1+ \mu_1 + \mu_2)+ \mu_1 + \mu_2}.
\end{equation}
The maximal violation of SNL is achieved by the
TWB state ($\mu_1=\mu_2=0$), while upon increasing the
intensity of at least one of the seeding fields the SNL is eventually
reached.
\subsection{Negativity}
The nonclassical behaviour of a set of light modes has been often
related to the negativity of the Glauber-Sudarshan P-function,
which, in turn, prevents the description of the systems as a
classical statistical ensemble. Here, in order to quantify
negativity in terms of the photon number distribution, we employ the
criterion introduced by Lee \cite{Lee1,Lee2}, which represents the
two-mode generalization of the Mandel's criterion of nonclassicality
\cite{Man79} for single-mode beams and, in turn, implies the
negativity of the P-function. According to Lee \cite{Lee1, Lee2}, a
bipartite system shows nonclassical behaviour if the inequality
\begin{equation} \label{lee}
\langle n_{1}(n_{1}-1)\rangle + \langle n_{2}(n_{2}-1)\rangle - 2
\langle n_{1} n_{2}\rangle < 0
\end{equation}
is satisfied. For the PDC output state, the
condition in Eq. (\ref{lee}) corresponds to $ \mu_1^2 + \mu_2^2 -
\mu_1  \mu_2<  \mu_\kappa (1+ \mu_1 + \mu_2)$.  As we did in the
case of intensity correlations, we define a parameter quantifying the
amount of negativity
\begin{equation} \label{Plee0}
\gamma_n= 1- \frac{\langle \Delta H ^{2}\rangle+ (\langle {n}_{1}
\rangle - \langle {n}_{2} \rangle)^2}{\langle {n}_{1} \rangle +
\langle {n}_{2} \rangle} .
\end{equation}
We have $0 < \gamma_n \leq 1$, with $\gamma_n=1$ corresponding to
maximum nonclassicality. For the PDC output state we
obtain
\begin{equation} \label{Plee}
\gamma_n =2 ~ \frac{ \mu_\kappa (1+ \mu_1 + \mu_2)- \mu_1^2 -
\mu_2^2+\mu_1 \mu_2}{2 ~ \mu_\kappa (1+ \mu_1 + \mu_2)+ \mu_1 +
\mu_2} .
\end{equation}
Again the most nonclassical state is the TWB state ($\mu_1=\mu_2=0$), whereas
by increasing the intensity of at least one of the seeding field,
the positive P-function region is eventually reached.
\subsection{Entanglement}\label{entsep}
The PDC process provides pairwise entanglement in the two modes. In
the spontaneous process the output state is entangled for any value
of $\mu_\kappa \neq 0$, whereas for thermally seeded PDC the degree
of entanglement crucially depends on the intensity of the seeding fields
\cite{DBA07}. For a bipartite Gaussian state,
entanglement is
equivalent to the positivity under partial transposition (PPT)
condition \cite{simon00}, which may be written in terms of the
smallest partially transposed symplectic eigenvalue. Thus, seeded PDC produces an entangled output state if and only if
\cite{DBA07}
\begin{equation}\label{autovalore}
\mu_{1} \mu_{2}- \mu_\kappa (1+\mu_{1} + \mu_{2})\geq 0.
\end{equation}
Remarkably, entanglement properties of the state $\bmr$ can be
verified by intensity measurements independently performed on
the two modes. In fact, with an ideal
detection system, the inequality
\begin{equation} \label{PPTn}
 \langle \Delta H^{2}\rangle - (\langle {n}_{1} \rangle - \langle
{n}_{2} \rangle)^2 \leq \langle {n}_{1} \rangle + \langle {n}_{2}
\rangle.
\end{equation}
reproduces exactly the entanglement condition in Eq. (\ref{autovalore}).
Therefore, the
amount of the violation of the separability boundary may be quantified
by means of the parameter
\begin{equation} \label{Pse1}
\gamma_e= 1- \frac{\langle \Delta H^{2}\rangle- (\langle {n}_{1}
\rangle - \langle {n}_{2} \rangle)^2}{\langle {n}_{1} \rangle +
\langle {n}_{2} \rangle}.
\end{equation}
$\gamma_e= 0$ corresponds to the boundary between separable and
entangled states. For the PDC output $\bmr$ we obtain
\begin{equation} \label{Pse2}
\gamma_e=2 ~ \frac{ \mu_\kappa (1+ \mu_1 + \mu_2)- \mu_1 \mu_2}{2 ~
\mu_\kappa (1+ \mu_1 + \mu_2)+ \mu_1 + \mu_2}.
\end{equation}
Maximally entangled states ($\gamma_e=1$) thus correspond to the
TWB ($\mu_1=\mu_2=0$), whereas entanglement is degraded in the presence of
thermal seeds. Notice, however, that if one of the two modes at the input is the
vacuum, the state is always entangled irrespective of the intensity
of the other seeds.
\par
\begin{figure}[h]
\begin{center}
\includegraphics[width=0.75\textwidth]{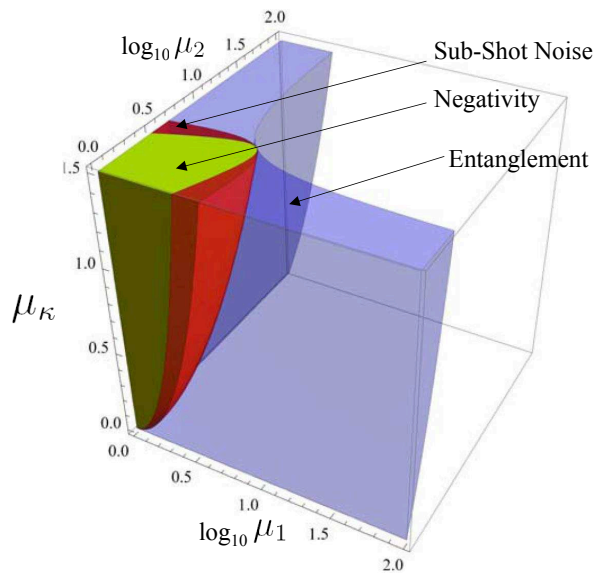}
\caption{(Color online) Nonclassicality regions, {\em i.e.} regions
for which $0<\gamma<1$, in terms  of $\mu_{1}$, $\mu_{2}$, and
$\mu_\kappa$ for the three $\gamma$ parameters introduced in the
text. As it is apparent from the plot, a hierarchy of regions and
bounds appears. The wider region (red$+$green$+$blue) identifies the
values of the $\mu$'s leading to an entangled output
($0<\gamma_e<1$) from the PDC. The intermediate (red$+$green)
corresponds to nonclassical intensity correlations ($0<\gamma_c<1$),
whereas the narrower internal region (green) is for negativity
($0<\gamma_n<1$). }\label{fig1} \end{center}
\end{figure}

In Fig. \ref{fig1} we show the nonclassicality regions in terms of
the seeding, $\mu_j$, $j=1,2$, and downconversion, $\mu_k$, mean
photon numbers, {\em i.e.} the triples $(\mu_1, \mu_2, \mu_\kappa)$
for which the parameters $\gamma$ lie in the interval
$0<\gamma\leq1$. As it is apparent from the plot, a hierarchy of
nonclassicality concepts and thresholds naturally appears. The most
stringent criterion of nonclassicality corresponds to negativity,
followed by sub-shot-noise intensity correlations and then by
entanglement.
\par
\begin{figure}[h]
\includegraphics[width=8cm]{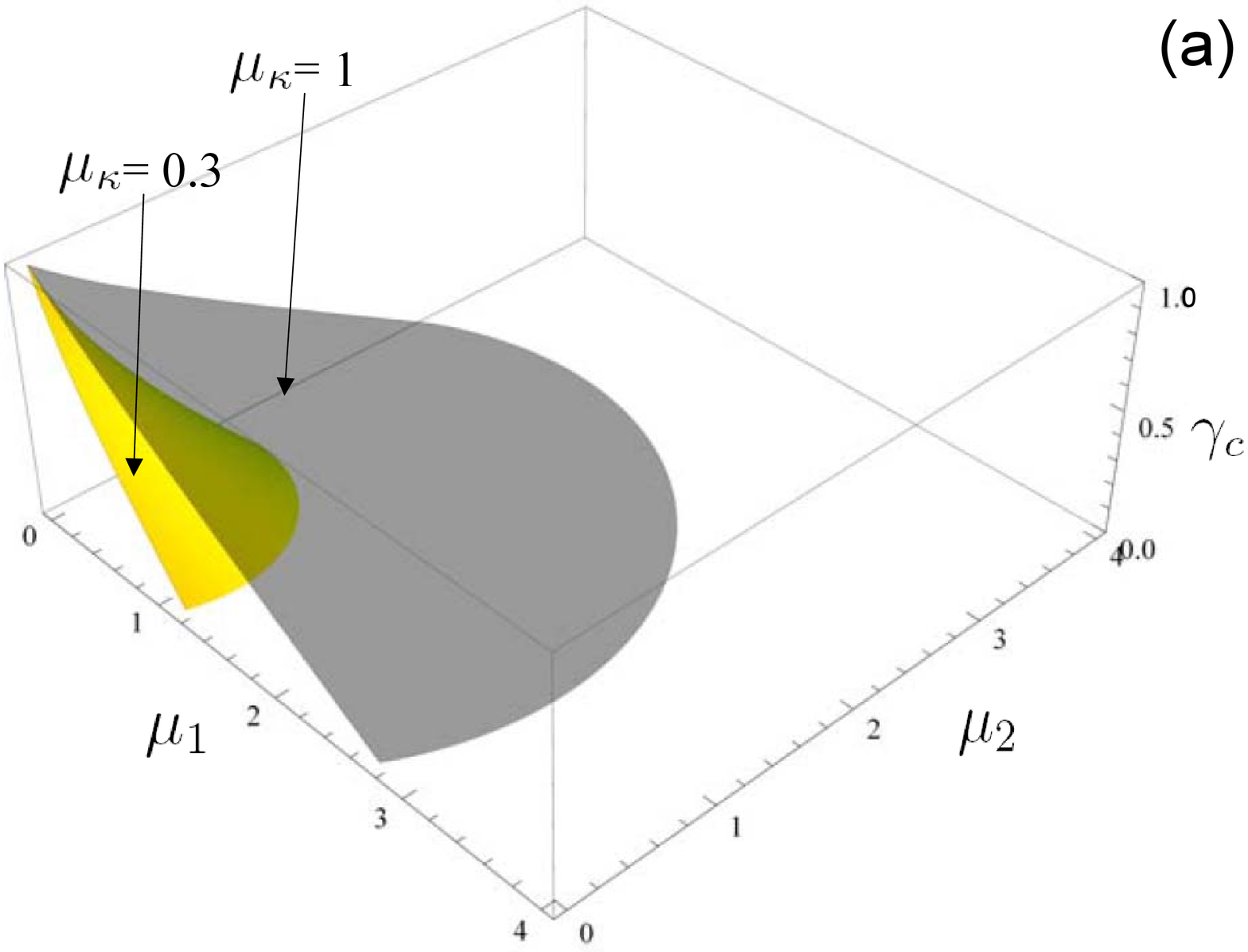}
\includegraphics[width=8cm]{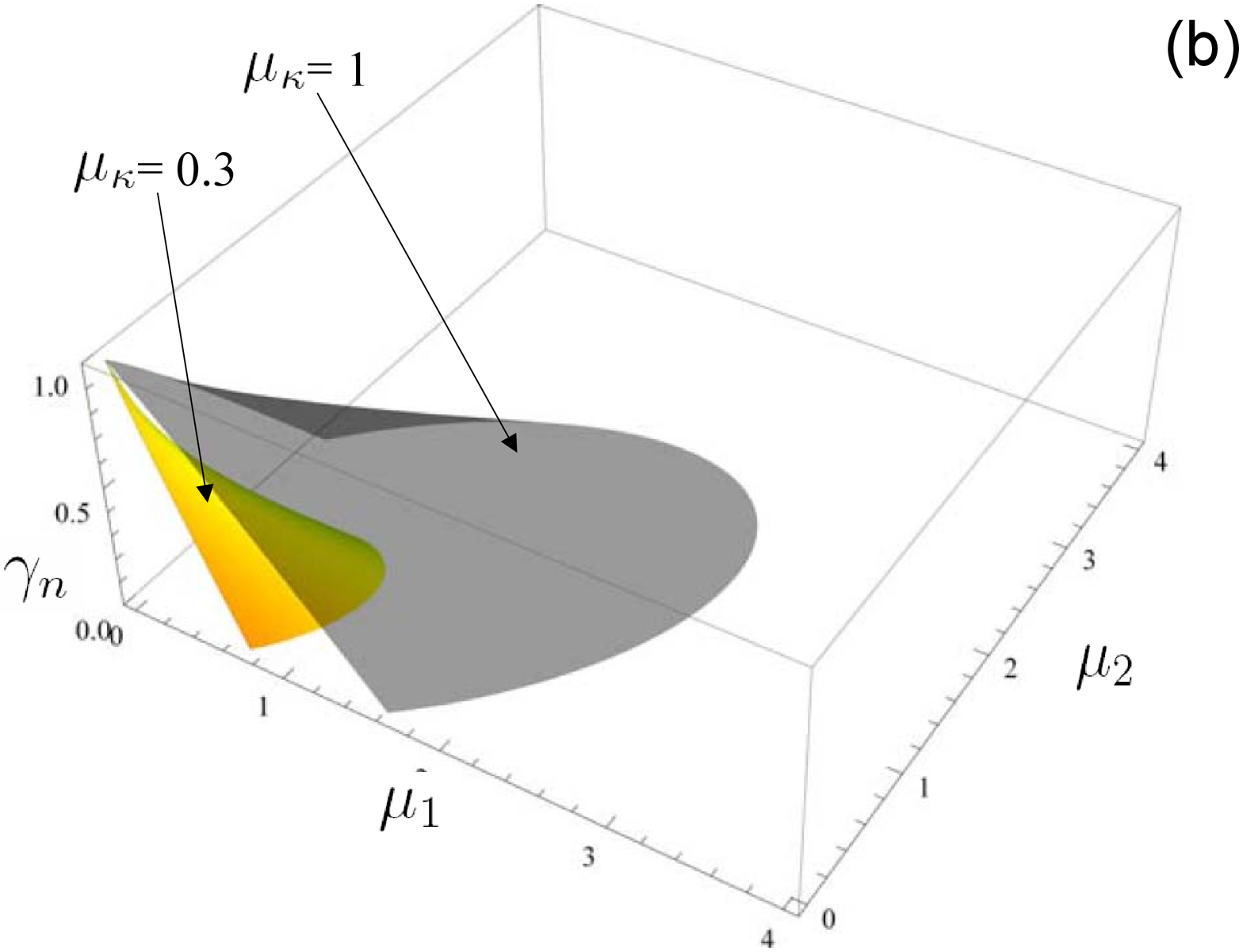}
\includegraphics[width=8cm]{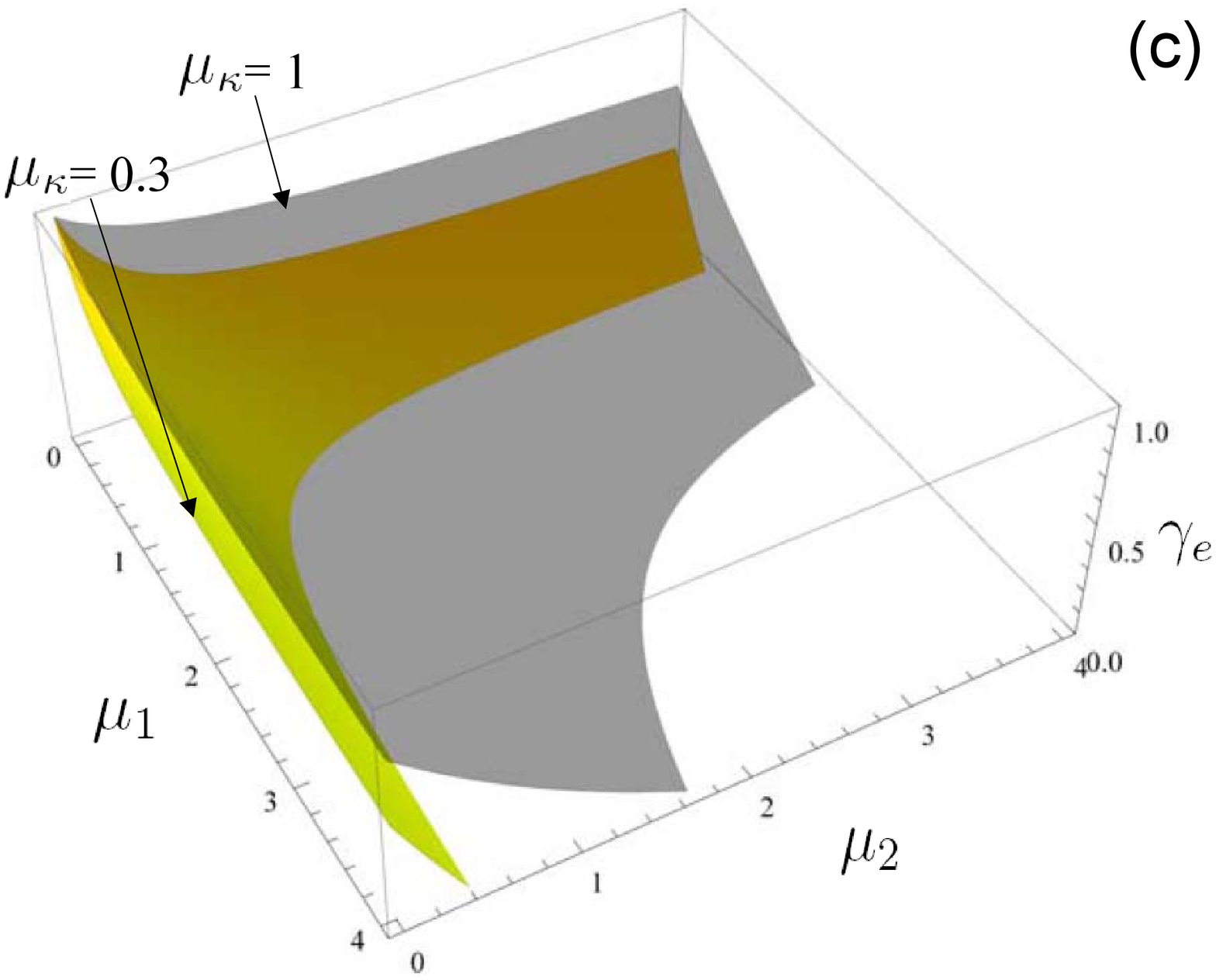}
\caption{(Color online) Nonclassicality parameters $\gamma_c$,
$\gamma_n$ and $\gamma_e$ evaluated for $\mu_\kappa=0.3$ (yellow)
and $\mu_\kappa=1$ (black) as a function of $\mu_{1}$ and
$\mu_{2}$.}\label{fig2}
\end{figure}

We can express the thresholds for the appearance of nonclassicality as conditions on
the mean number of photons resulting from the downconversion process
\begin{eqnarray}
{\gamma_n=0} &\rightarrow& \mu_\kappa^{n}  =
\frac{\mu_1^2+\mu_2^2-\mu_1\mu_2}{1+\mu_1+\mu_2} \quad
\\
{\gamma_c=0} &\rightarrow& \mu_\kappa^{c}  =
\frac{\mu_1^2+\mu_2^2}{2 (1+\mu_1+\mu_2)}
\\
{\gamma_e=0} &\rightarrow& \mu_\kappa^{e}  =
\frac{\mu_1\mu_2}{1+\mu_1+\mu_2} \quad
\end{eqnarray}
In other words, being negative-nonclassical is a sufficient
condition to have sub-shot-noise intensity correlation. Moreover, either of
the two (negativity and sub-shot-noise) is a sufficient condition
for entanglement, {\em i.e.} $\mu_\kappa^{c}>\mu_\kappa^{n}>\mu_\kappa^{e}$
for any value of $\mu_1$ and $\mu_2$. Remarkably, the three nonclassicality
conditions collapse into a single one when the seeding intensities are
equal $\mu_{1}=\mu_{2}$ and differ by terms up to the second order in
$|\mu_1 -\mu_2|$ in the neighbourhood of this condition.
It is already evident in Fig. \ref{fig1}, as well as in Fig. \ref{fig2},
where we show the three parameters as
function of the seeding intensities for different values of
$\mu_\kappa$, that the stronger is
the spontaneous PDC (large $\mu_\kappa$) the larger is the number
of thermal photons that can be injected while preserving negativity
and hence sub-shot-noise correlations and entanglement.
\section{Effect of losses}
In order to see whether the nonclassicality thresholds identified in
the previous section may be investigated experimentally, one should
take into account losses occurring during propagation, which
generally degrade quantum features, and non-unit quantum efficiency
in the detection stage, which may prevent the demonstration of
nonclassicality. The two mechanisms may be subsumed by an overall
loss factor $\tau$ \cite{R20, R21} using a beam splitter model
\cite{R22,rob} in which the signal enters one port and the second
one is left unexcited. Upon tracing out the second mode we describe
the loss of photons during the propagation and the detection stage.
In the following we assume equal transmission factor for the two
channels and evaluate the nonclassicality parameters in the presence
of losses.
\par
\begin{figure}[h]
\includegraphics[width=8cm]{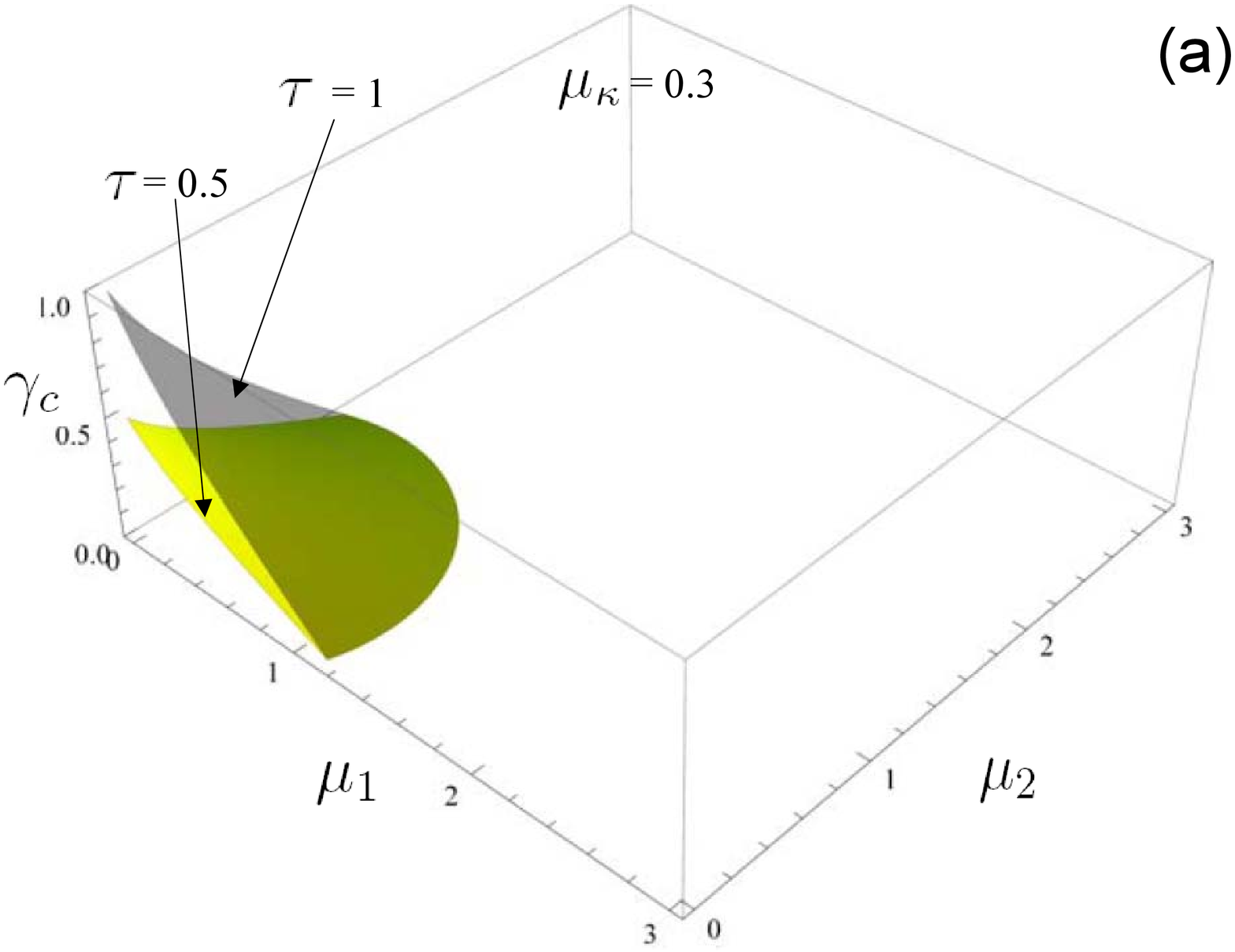}
\includegraphics[width=8cm]{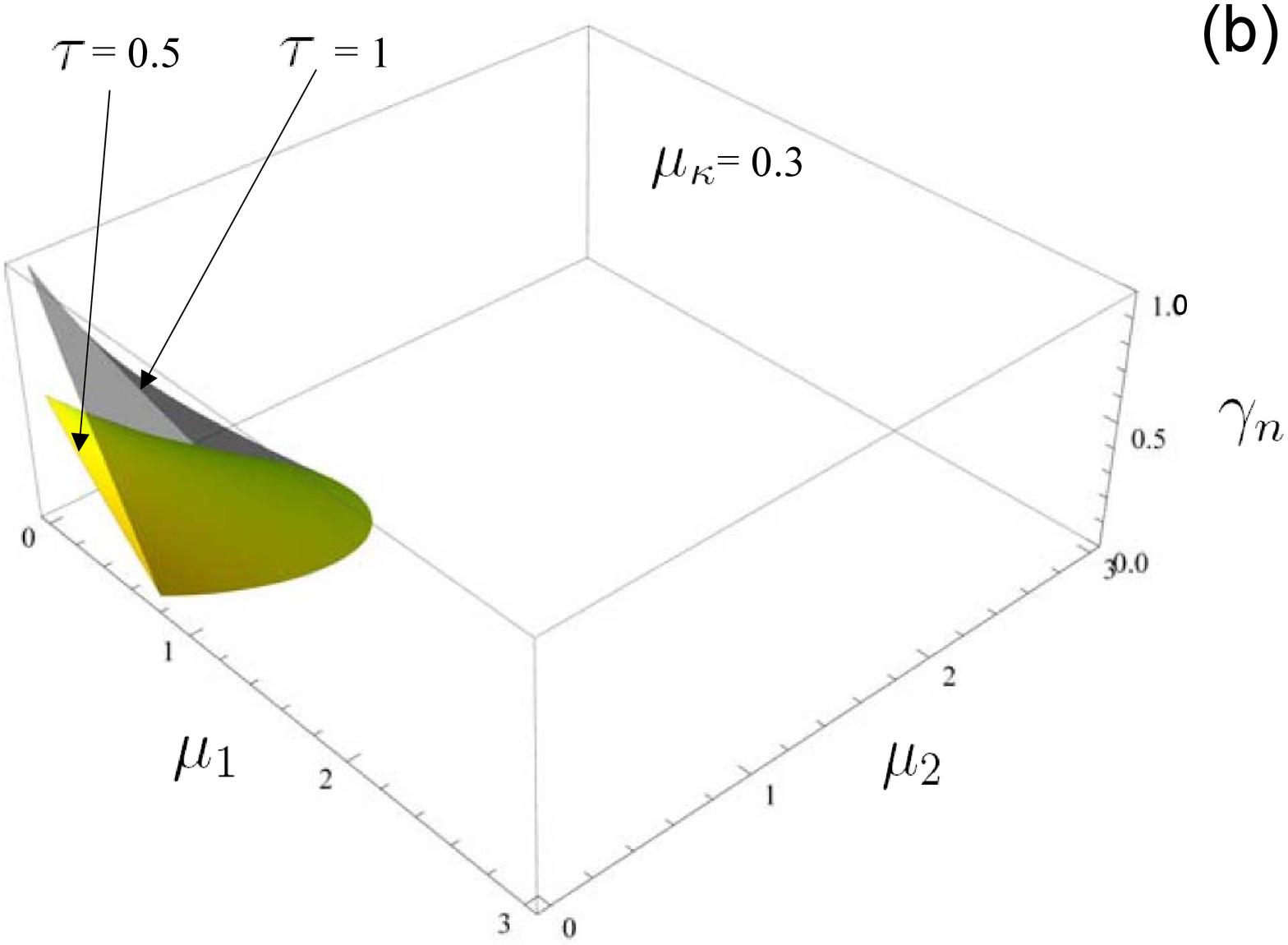}
\includegraphics[width=8cm]{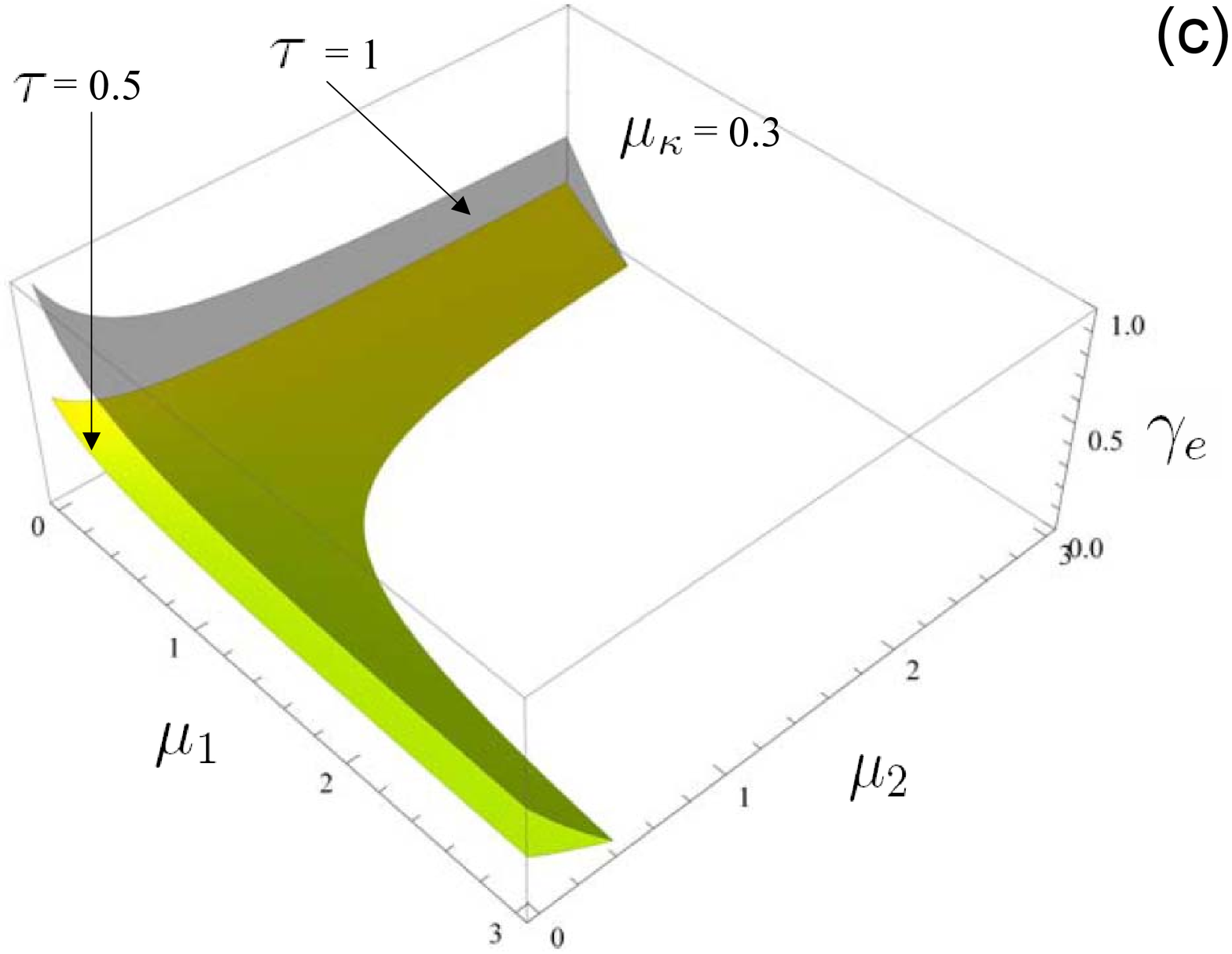}
\caption{(Color online) Nonclassicality parameters $\gamma_c$,
$\gamma_n$ and $\gamma_e$ evaluated for $\tau=0.5$ (yellow) and in
the absence of losses ($\tau=1$, black) as a function of $\mu_{1}$
and $\mu_{2}$, with $\mu_\kappa=0.3$.}\label{fig3}
\end{figure}

 Upon assuming that dark counts have been already subtracted,
the positive operator-valued measure (POVM) of each detector is
given by a Bernoullian convolution of the ideal number operator
spectral measure. The moments of the distribution are evaluated by
means of the operators
\begin{equation}
N_j(\tau,p)  = \sum_{n=0}^\infty (1-\tau)^n \:G_{j}(p,n)\:
|n\rangle\langle n |
\end{equation}
where $G_j(p,n)= \sum_{m=0}^{n} {n\choose m}
\left(\frac{\tau}{1-\tau}\right)^{m}\!\! m^p$. Of course, since
$N_j(\tau,p) $ are operatorial moments of a POVM, in general we have
$N_j(\tau, p) \neq N_j(\tau,1)^p$, with the first two moments given
by
\begin{eqnarray}
 N_j(\tau,1)  =\tau  n_j  \\
N_j(\tau,2) =\tau^2  n_j^2 + \tau (1-\tau)  n_j \:.
\end{eqnarray}
Upon inserting the above expressions in the nonclassicality
parameters ({\em i.e.} replacing $n_j$ and $n_j^2$ by $N_j(\tau,1)$
and $ N_j(\tau,2)$ respectively), we obtain that, for all of them,
the inclusion of losses results in a simple re-scaling
\begin{equation}
\gamma_i (\tau) = \tau \, \gamma_i (\tau = 1) \qquad i=c,n,e\:.
\end{equation}
In other words, the effect of losses is that of decreasing the
amount of nonclassicality, whereas the thresholds for the
quantum-classical transitions are left unaffected. This also means
that the twin-beam still corresponds to the maximal violation of
classicality condition independently of the kind of nonclassicality
parameter we are taking into account. These are shown in Fig.
\ref{fig3}, where the parameters $\gamma$ for $\tau=0.5$ are
compared with those in ideal condition for a fixed value of the PDC
gain.
\section{The multimode case}
The (quantum) correlations introduced by the PDC process are
intrinsically pairwise and thus no qualitative differences should be
expected when considering the multimode case. On the other hand, the
expression of the parameters $\gamma$ does depend on the number of
modes and thus it is worth explicitly addressing the multimode case
\cite{DBA07}. Besides, from the experimental point of view, this is
a situation often encountered in travelling-wave PDC pumped by
pulsed lasers.
\par
The evolution operator for the
multimode case can be rewritten in terms of the operators
${S}_{\xi}=(\kappa_{\xi} {a}_{1,\xi} {a}_{2,\xi} + h.c.)$ as
${U}_M=\bigotimes_{\xi=1}^M\: e^{i {S}_{\xi}}$ thus emphasizing the pairwise
structure.
In our analysis we focus on the case in which all the modes are
seeded with uncorrelated multi-mode thermal fields with
$\mu_{j,\xi}$ mean photon number per mode
\begin{eqnarray}\label{rhoin}
{\bmn_{M}}=\bigotimes_{\xi=1}^M\left(
{\nu}_{1,\xi}\otimes {\nu}_{2,\xi} \right)\qquad
{\nu}_{j, \xi}=\sum_{n=0}^{\infty}p_{j,\xi}(n)~
|n\rangle_{j,\xi}{}_{\xi,j}\langle n |\nonumber\:,
\end{eqnarray}
where $j=1,2$. The density matrix at the output is thus given by
\begin{equation}\label{rhooutq}
\bmrm = U_M \bmn_{M} U^{\dagger}_M= \bigotimes_{\xi}\left[ e^{i
{S}_{\xi}}\: \left(\nu_{1,\xi} \otimes \nu_{2,\xi} \right) \: e^{-i
{S}_{\xi}} \right] \:,
\end{equation}
and the calculation for each pair of coupled modes is completely
analogous to that performed in the first Section (see also
\cite{DBA07}). The Heisenberg evolution of modes is
\begin{equation}
{A}_{j,\xi}={U}^{\dag}{a}_{j,\xi}{U}=\alpha_{\xi}{a}_{j,\xi}+ e^{i
\varphi_{\xi}}\beta_{\xi}{a}_{j',\xi}^{\dag} ~~ (j,j'=1,2 , ~j \neq
j') \label{aout2}
\end{equation}
where $\alpha_{\xi}=\cosh|\kappa_{\xi}|$ and $\beta_{\xi}=
\sinh|\kappa_{\xi}|$. The spontaneous PDC energy for each pair
of modes is $\mu_{\kappa,\xi}=\sinh^2|\kappa_{\xi}|$.
In this case, the number of photons measured in each arm is
$n_{j}=\sum_{\xi} n_{j, \xi}$, with $n_{j, \xi}=a_{j, \xi}^{\dagger}
a_{j, \xi}$ ($j=1,2$). The quantities relevant to our
analysis are the mean photon values $\langle n_{j} \rangle=
\sum_{\xi} \langle n_{j, \xi} \rangle$ and the variances of the
difference photocurrent $H=\sum_\xi H_\xi$, $H_{\xi}=n_{1, \xi}
- n_{2, \xi}$. Since correlations are only pairwise we have
$ \langle n_{j, \xi} n_{j', \xi'} \rangle = \langle
n_{j, \xi} \rangle \langle n_{j', \xi'} \rangle $ when $ \xi \neq
\xi'$ and thus
\begin{equation} \label{mm1}
\langle \Delta H^{2}\rangle = \sum_{\xi} \langle \Delta H_{
\xi}^{2}\rangle .
\end{equation}
Using this result, the extension to the multimode case for intensity
correlations is straightforward, and the violation of the SNL in Eq.
(\ref{ssn}) can be rewritten as
\begin{equation}\label{modesSNL}
\sum_{\xi}   \left( \langle \Delta H_{ \xi}^{2}\rangle -\langle
{n}_{1, \xi} \rangle - \langle {n}_{2, \xi} \rangle \right) <0.
\end{equation}
If we assume that each mode of the seeding thermal fields in the $j$
arm ($j=1,2$) has the same mean photon number, $\mu_{j,\xi}=
\mu_{j}$, and that the parametric gain is the same for each pair of
coupled mode, $\mu_{\kappa,\xi}= \mu_\kappa$, the condition for the
violation of the SNL in the multimode case is the same as for the
single-mode seeds. The same is true in presence of losses, upon
assuming equal transmission factor $\tau$ for the modes, as it can
also easily seen by inspecting Eq. (\ref{modesSNL}). On the other
hand, for the negativity, as expressed by Eq. (\ref{Plee0}), the
extension to the multimode case is not possible, since its
derivation is explicitly based on the assumption of a single pair of
downconverted modes \cite{Lee1,Lee2}.
\par
Finally, the separability/entanglement condition for the multimode
thermally seeded PDC has already been analyzed \cite{DBA07}, and it
has been demonstrated that the separability properties of state
$\bmrm$ may be checked by intensity measurements on the two arms,
though not for a generic multimode field. An interesting case is
when each mode of the seeding thermal fields in the $j$ arm
($j=1,2$) has the same mean number of photons, $\mu_{j,\xi}=
\mu_{j}$, and the parametric gain is the same for each pair of
coupled modes, $\mu_{\kappa,\xi}= \mu_\kappa$. In this case, the
entanglement condition is still given by Eq. (\ref{autovalore}) and
it is possible to reveal entanglement of the state $\rho_{M}$ by
means of direct photon counting measurements on $1$ and $2$ arms
exploiting the inequality
\begin{equation} \label{PPTnMM}
 \langle \Delta
H^{2}\rangle - \frac{(\langle {n}_{1} \rangle - \langle {n}_{2}
\rangle)^2}{\mathcal{N}} \leq \langle {n}_{1} \rangle + \langle
{n}_{2} \rangle.
\end{equation}
which is almost equal to Eq. (\ref{PPTn}) except for the second term
where the number of modes $\mathcal{N}$ appears. In fact, by
starting from Eq. (\ref{modesSNL}) and substituting the multimode
expression of $n_1$ and $n_2$, it can easily be proved that Eq.
(\ref{PPTnMM}) leads to the entanglement condition in Eq.
(\ref{autovalore}). As it has already been demonstrated \cite{DBA07}
that the boundary between separability and entanglement is not
modified by presence of losses, it is straightforward to prove that
Eq. (\ref{PPTnMM}) still holds.
\section{Conclusions and outlooks}
\label{s:out} In this paper we have addressed the quantum-classical
transition for the radiation field in the continuous variable
regime. We have analyzed in detail the pair of conjugated field
modes obtained from parametric-downconversion and explicitly
evaluated intensity correlations, negativity and entanglement for
the system seeded by radiation in a thermal state. Our results have
shown that a hierarchy of nonclassicality thresholds naturally
emerges in terms of thermal and downconversion photon number and
that the transition from quantum to classical regime may be tuned by
controlling the seed intensities. The quantum/classical thresholds
derived in this paper have two features that make them appealing for
an experimental verification: i) they are not affected by losses,
which only modify the amount of violation; ii) they can be verified
by intensity measurements, without phase-dependent measurements and
full state reconstruction.  According to Fig. \ref{fig1}, in order
to appreciate the differences among the criteria discussed above,
the fields should contain a non-negligible number of photons coming
both from the PDC process and from the seeds. We plan to generate
such states by frequency-degenerate, noncollinear, travelling-wave
PDC pumped by a high energy pulsed laser \cite{subshot07}. In the
experiment we should take advantage of the fact that $\mu_\kappa$
can be reasonably high and bring the nonclassicality parameters to
interesting regions. Tens of photons are expected from the process
that may be measured by a pair of linear photodetectors with
internal gain (photomultipliers or hybrid photodetectors) as
described in \cite{ASL08}. Besides, as an alternative to
conventional crystals, a periodically-poled-non-linear waveguide
medium and CW laser may be employed, aiming at the production of
inherently single-mode (frequency) non-degenerate PDC light.
Extension to the tripartite case \cite{tr1,tr2,tr3, R31, R32, R33,
R34, R35, R36, R37, R38} is also in progress and results will be
reported elsewhere.
\section*{Acknowledgments}
M.G.A.P. thanks S. Olivares and M. Genoni for useful discussions.
M.G. and I.P. D. thank M. Chekhova for useful discussions.  This
work has been supported by C.N.I.S.M., Regione Piemonte (E14),
SanPaolo Foundation, Lagrange Project CRT Foundation.

\end{document}